\documentclass{article} 
\usepackage{iclr2025_conference,times}

\usepackage{amsmath,amsfonts,bm}

\def\eqref#1{equation~\ref{#1}}

\def\1{\bm{1}}

\DeclareMathAlphabet{\mathsfit}{\encodingdefault}{\sfdefault}{m}{sl}
\SetMathAlphabet{\mathsfit}{bold}{\encodingdefault}{\sfdefault}{bx}{n}

\usepackage{hyperref}
\usepackage{url}
\usepackage{booktabs}
\usepackage{graphicx}          
\usepackage[section]{placeins} 
\usepackage[export]{adjustbox} 

\newlength{\StdFigH}
\title{A Biosecurity Agent \\for Lifecycle LLM Biosecurity Alignment}

\author{
  Meiyin Meng\\
  Imperial College London\\
  \texttt{mm3222@ic.ac.uk}
  \And
  Zaixi Zhang\thanks{Corresponding author.}\\
  Princeton University\\
  \texttt{zz8680@princeton.edu}
}

\iclrfinalcopy 

\begin{document}
\maketitle
\begin{abstract}
Large language models (LLMs) are increasingly integrated into biomedical research workflows—from literature triage and hypothesis generation to experimental design—yet this expanded utility also heightens dual-use concerns, including the potential misuse for guiding toxic compound synthesis. In response, this study shows a Biosecurity Agent that comprises four coordinated modes across the model lifecycle: dataset sanitization, preference alignment, runtime guardrails, and automated red teaming. For dataset sanitization (\textbf{Mode~1}), evaluation is conducted on CORD-19, a COVID-19 Open Research Dataset of coronavirus-related scholarly articles. We define three sanitization tiers—L1 (compact, high-precision), L2 (human-curated biosafety terms), and L3 (comprehensive union)—with removal rates rising from 0.46\% to 70.40\%, illustrating the safety–utility trade-off. For preference alignment (\textbf{Mode~2}), DPO with LoRA adapters internalizes refusals and safe completions, reducing end‑to‑end attack success rate (ASR) from 59.7\% to 3.0\%. At inference (\textbf{Mode~3}), runtime guardrails across L1–L3 show the expected security–usability trade-off: L2 achieves the best balance (F1 = 0.720, precision = 0.900, recall = 0.600, FPR = 0.067), while L3 offers stronger jailbreak resistance at the cost of higher false positives.
Under continuous automated red‑teaming (\textbf{Mode~4}), no successful jailbreaks are observed under the tested protocol. Taken together, our biosecurity agent offers an auditable, lifecycle-aligned framework that reduces attack success while preserving benign utility, providing safeguards for the use of LLMs in scientific research and setting a precedent for future agent-level security protections.
\end{abstract}

\section{Introduction}
Large language models enable rapid literature triage, drafting, and knowledge access in the life sciences \citep{liang2022helm,openai2024gpt4}. This capability also entails \emph{dual-use} risk when unsafe instructions or tacit know-how are elicited \cite{wang2025call}. Recent taxonomies and risk surveys characterize such hazards and recommend layered safeguards with continuous evaluation \citep{weidinger2022taxonomy,weidinger2021risks,shevlane2023extremerisk}. Governance frameworks emphasize pre-deployment assessment, ongoing monitoring, and domain-aware controls for high-stakes applications, as reflected in the U.S.\ Executive Order~14110, the EU AI Act, and the NIST AI Risk Management Framework \citep{whitehouse2023eo14110,eurlex2024aiact,nist2023airmf}. Foundational work on modern AI and LLMs further motivates safety alignment in sensitive domains \citep{goodfellow2016deep,Bengio+chapter2007,Hinton06,openai2024gpt4}.

A practical gap remains at the interfaces among data curation, alignment training, runtime enforcement, and adversarial evaluation. Evidence from benchmarks and in-the-wild studies indicates that defenses deployed in isolation leave these interfaces exposed, allowing jailbreak prompts to bypass safeguards or exploit blind spots \citep{chao2024jailbreakbench,mazeika2024harmbench,li2024wildjailbreak,liu2024autodan,zou2023universal, zhang2025genebreaker, fan2025safeprotein}. The scope of this work is limited to \textbf{text-only LLMs} for natural-language assistance in biology. Sequence-level generative models (e.g., for DNA or proteins) and multimodal lab-control systems are outside the scope of this study.

\begin{figure*}[!t]
  \centering
  \includegraphics[width=\linewidth,keepaspectratio]{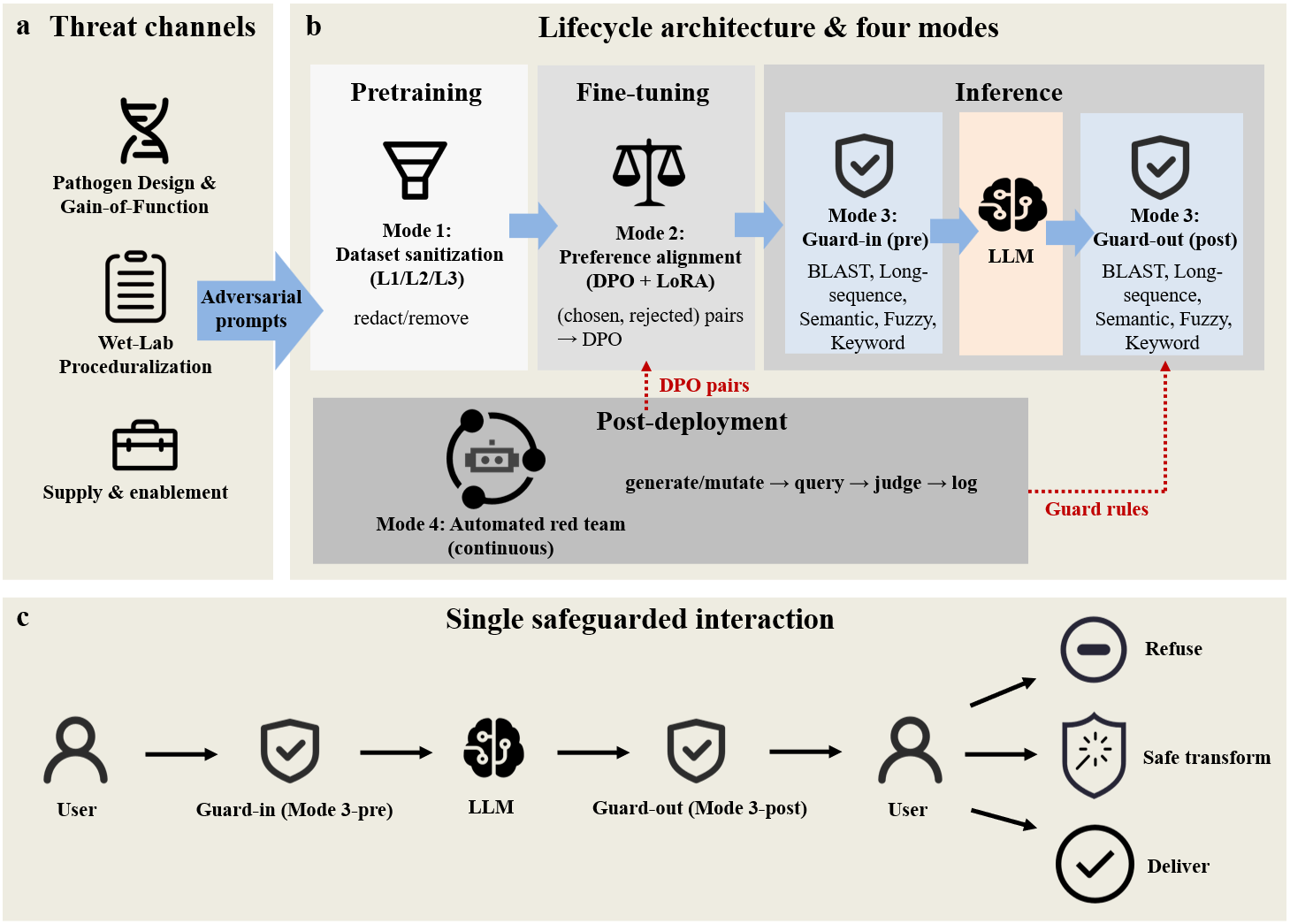}
  \caption{\textbf{Overview of the defence-in-depth Biosecurity Agent.}
  Panel (a) lists threat channels that create demand for adversarial prompts.
  Panel (b) shows the lifecycle architecture with four modes.
  Mode~1 performs dataset sanitization with keyword tiers L1/L2/L3.
  Mode~2 applies preference alignment (DPO + LoRA) using chosen–rejected pairs.
  Mode~3 enforces runtime guardrails at input and at output by combining BLAST, long-sequence, semantic, fuzzy, and keyword checks.
  Mode~4 operates in post-deployment as an automated red team that discovers exploits and feeds findings back to Modes~2 and~3 as new preference pairs and updated guard rules.
  Panel (c) illustrates a single safeguarded interaction that follows Eq.~\eqref{eq:pipeline}.
  The deployment target is an attack success rate below five percent.}
  \label{fig:overview}
\end{figure*}

As summarised in Fig.~\ref{fig:overview}, a \emph{defence-in-depth} toolkit for LLM biosecurity alignment is operationalised as a tool-orchestrated agent covering all stages of the model lifecycle. Training data are curated, model behaviour is aligned, inference is gated, and residual failures are discovered and fed back into the process. Mode~1 performs dataset sanitisation with tiered keyword filtering to remove or redact risky content, informed by biosecurity guidance and screening practice \citep{who2022lifesciences,nsabb2023framework,igsc2024hsp}. Mode~2 applies preference alignment using Direct Preference Optimization (DPO) with LoRA adapters to internalise refusals and safe completions \citep{rafailov2023dpo,hu2022lora,bai2022constitutional,ouyang2022instructgpt}. Mode~3 enforces runtime guardrails with pre- and post-generation checks that aggregate multiple biology-aware signals. LLM-based safety classifiers and guardrail stacks provide programmable policy enforcement \citep{llamaguard2023,shankar2024guardrails}. Robust smoothing further complements these mechanisms \citep{robey2023smoothllm}. Mode~4 conducts automated red teaming that continually discovers exploits and updates Modes~2 and~3. Moreover, public benchmarks and autonomous attackers support standardized evaluation and continuous discovery \citep{chao2024jailbreakbench,mazeika2024harmbench,zhou2025autoredteamer,li2024wildjailbreak,liu2024autodan}.

Evidence on curated corpora and stress tests supports this design. On CORD-19, Mode~1 yields a monotonic removal curve at three biosecurity filtering levels with 0.46\% of records removed at L1, 20.87\% at L2, and 70.40\% at L3. Under preference alignment, Mode~2 reduces end-to-end attack success (ASR) from 59.7\% (95\% CI 55.6–63.7) to 3.0\% (1.0–5.0). At inference time, Mode~3 exhibits a clear security–usability trade-off: a balanced operating point attains strong precision and recall with a low false-positive rate (FPR), whereas stricter settings further reduce jailbreak success at the cost of additional false positives. With continuous automated red teaming, Mode~4 improves post-alignment robustness: precision and recall increase, FPR decreases, and protection shifts upstream toward the pre-guard stage.

\section{Related Work}

\paragraph{LLM safety landscape and scope.}
Prior research on LLM safety has addressed multiple layers of the model lifecycle, including dataset curation, training-time alignment, inference-time safeguards, and adversarial evaluation. Benchmarks and audits consistently show that each layer in isolation is insufficient, motivating lifecycle approaches with continuous monitoring and explicit operating-point control \citep{openai2024gpt4,liang2022helm,chao2024jailbreakbench,mazeika2024harmbench}. Existing efforts span dataset filtering pipelines, alignment techniques such as RLHF and DPO, runtime moderation and guardrails, as well as red-teaming frameworks. The present work builds on this landscape by composing these elements into a unified, defense-in-depth framework—dataset sanitization, preference alignment, runtime guardrails, and automated red teaming—targeted at text-only LLMs as a first step toward secure deployment in scientific domains.

\paragraph{Dataset-level filtering and preprocessing.}
Unsafe behaviour frequently traces to unsafe training data. Toxic passages, hazardous instructions, and tacit procedural cues can lead to undesirable behaviours if left unfiltered. Dataset-level sanitisation therefore removes or redacts risky content before training, guided by biosecurity norms and screening protocols \citep{who2022lifesciences,nsabb2023framework,igsc2024hsp}. Data risks also include memorisation and privacy leakage, which have been demonstrated in large models \citep{carlini2021extracting,nasr2023scalable}. Evaluation corpora such as RealToxicityPrompts support measurement of degenerate toxic generation \citep{gehman2020realtoxicityprompts}.

\paragraph{Alignment during training.}
Alignment after pretraining reduces unsafe completions and improves helpfulness. Instruction-following via RLHF optimises models toward human preferences \citep{ouyang2022instructgpt}. DPO reframes preference alignment as a supervised objective and simplifies training without a reward model \citep{rafailov2023dpo}. LoRA enables parameter-efficient fine-tuning by injecting low-rank adapters into transformer layers \citep{hu2022lora}. Constitutional-style feedback guides self-critique against written principles and improves harmlessness with limited annotation burden \citep{bai2022constitutional}. Decoding-time control methods such as DExperts and PPLM complement alignment by steering generations away from unsafe modes \citep{liu2021dexperts,dathathri2020pplm}.

\paragraph{Inference-time safeguards.}
Adversarial prompts at deployment continue to elicit unsafe outputs. Runtime safeguards gate inputs and outputs with lightweight checks and can transform prompts to neutralize hidden triggers. Threat surveys catalog common vectors and mitigations for prompt-injection and jailbreaking \citep{liu2023promptinjection}. LLM-based safety classifiers and guardrail stacks provide programmable policy enforcement \citep{llamaguard2023,shankar2024guardrails}. Robust smoothing randomizes inputs or decoding and aggregates outcomes to detect adversarial sensitivity, with large reductions in jailbreak success reported \citep{robey2023smoothllm}.

\paragraph{Automated red teaming and standardised evaluation.}
Manual red teaming reveals failure modes but lacks coverage at scale. Public benchmarks support systematic evaluation across diverse harms and policy regimes \citep{chao2024jailbreakbench,mazeika2024harmbench}. Automated attackers further extend coverage: universal and transferable jailbreaks, optimization-based prompt generation, and in-the-wild analyses reveal broad attack surfaces \citep{zou2023universal,liu2024autodan,li2024wildjailbreak}. Autonomous red-teaming agents make evaluation continuous by discovering new attacks and integrating them over time \citep{zhou2025autoredteamer}. Transfer-focused evaluations also measure cross-model robustness \citep{jain2024artprompt}. These instruments motivate continuous hardening of defenses throughout the model lifecycle.

\section{Methods}
\label{sec:method}

\paragraph{System overview.}
All components are implemented as a tool-orchestrated \emph{Biosecurity Agent} using standard LLM planning/execution patterns \citep{yao2023react,khattab2024dspy}. The Biosecurity Agent itself is built upon our self-evolving STELLA agent framework \cite{jin2025stella}, through which biosecurity tools are integrated and experiments are executed in a closed-loop fashion. Under the default setting, the base model is \textbf{Llama-3-8B-Instruct} \citep{dubey2024llama3}, an instruction-tuned LLM with 8 billion parameters. All model training and inference used the Hugging Face \texttt{transformers} library \citep{wolf2020transformers} for reproducibility. The toolkit spans four lifecycle-aligned modes: \textbf{dataset sanitisation (Mode~1)}, \textbf{preference alignment (Mode~2)}, \textbf{runtime guardrails (Mode~3)}, and \textbf{automated red teaming (Mode~4)}.  
At inference time, the full pipeline can be formalized as:
\begin{equation}
\label{eq:pipeline}
\hat y \;=\; G_{\text{post}}\!\Big(M_\theta\!\big(G_{\text{pre}}(x)\big)\Big)\,,
\end{equation}
where $x$ is the raw user query, $G_{\text{pre}}$ denotes the guard functions that sanitize or transform the input (e.g., removing unsafe instructions, enforcing input constraints), and $M_\theta$ is the underlying language model parameterized by $\theta$, which maps inputs to candidate outputs in distributional form. The raw output of the model is then filtered by $G_{\text{post}}$, which applies output-side safety checks and transformations (e.g., refusal policies, structured output validation). This composition ensures that safeguards are applied both before and after the core generative model. Equation~\ref{eq:pipeline} thus provides a unified abstraction of the pipeline and will serve as the basis for defining pre-block failure rates (guard intercepts before model execution) and end-to-end failure rates (leaks past both pre- and post-guards) in subsequent sections.

\subsection{Datasets and evaluation protocol}
\label{sec:data_eval}
For \textbf{Mode~1}, we employed the CORD-19 corpus \citep{wang2020cord19}, a large collection of scholarly articles on coronaviruses and related biomedical topics widely used for text mining and language model evaluation, to exercise dataset-level sanitization. 
For \textbf{Mode~2}, we curated a preference dataset of JSONL triples 
\{\texttt{prompt}, \texttt{chosen}, \texttt{rejected}\} to train with DPO \citep{rafailov2023dpo}, 
ensuring that all candidate texts passed both pre- and post-generation guard filters to exclude non-compliant samples. 
For \textbf{Mode~3}, a balanced set of 60 prompts (30 harmful, 30 safe) was used to characterize guardrail operating points 
across three keyword strictness tiers (L1\_custom, L2\_human, L3\_all). 
We define the \emph{safe-accept rate} as the proportion of safe prompts that receive an allowed, policy-compliant answer 
after the full pipeline in Eq.~\eqref{eq:pipeline}. 
For \textbf{Mode~4}, we implemented an automated attacker loop that iteratively generated and mutated prompts across rounds 
\citep{zhou2025autoredteamer}. 
Unless otherwise noted, the default configuration set the pre-guard to strict mode without BLAST 
and the post-guard to strict mode with BLAST enabled.

\subsection{Metrics}
\label{sec:metrics}
Let $\text{TP}$, $\text{FP}$, $\text{TN}$, and $\text{FN}$ denote confusion-matrix counts, 
with “positive” corresponding to a harmful prompt. 
We report standard classification metrics in the usual way:
\begin{equation}
\text{Precision}=\tfrac{\text{TP}}{\text{TP}+\text{FP}},\quad
\text{Recall}=\tfrac{\text{TP}}{\text{TP}+\text{FN}},\quad
\text{F1}=\tfrac{2\,\text{Precision}\cdot\text{Recall}}{\text{Precision}+\text{Recall}},\quad
\text{FPR}=\tfrac{\text{FP}}{\text{FP}+\text{TN}}\,.
\end{equation}
In addition to these standard metrics, we introduce two safety-specific metrics to our pipeline.  
\paragraph{Pre-jailbreak success rate (pre\_JSR).}  
This metric quantifies how often the pre-guard fails to block harmful inputs.  
\begin{equation}
\text{pre\_JSR} = 
\frac{\#\,\text{harmful prompts not blocked by } G_{\text{pre}}}
     {\#\,\text{harmful prompts}}\,.
\end{equation}

\paragraph{Attack success rate (ASR).}  
This metric measures the end-to-end failure rate, i.e., the fraction of harmful prompts that ultimately reach the user after passing through the full pipeline of pre-guard, model, and post-guard:  
\begin{equation}
\text{ASR} = 
\frac{\#\,\text{harmful prompts that reach the user after pipeline}}
     {\#\,\text{harmful prompts}}\,,
\end{equation}
where “pipeline” denotes $G_{\text{post}} \!\circ M_\theta \!\circ G_{\text{pre}}$.  

For all proportion estimates, we report 95\% confidence intervals \citep{brown2001interval}:
\begin{equation}
\label{eq:cp}
\text{CI}_{95\%}(k,n)=
\Bigl[\,\mathrm{Beta}^{-1}(0.025;\,k,\,n{-}k{+}1),\ 
       \mathrm{Beta}^{-1}(0.975;\,k{+}1,\,n{-}k)\,\Bigr].
\end{equation}

\subsection{Mode 1: Dataset sanitisation}
\label{sec:mode1}
Dataset-level filtering was performed offline by a utility function \texttt{sanitize\_bio\_dataset}. All textual fields in each record were scanned and any match against the tiered keyword lists (L1\_custom, L2\_human, L3\_all) triggered redaction or removal of the content. 

\subsection{Mode 2: Safety alignment via DPO}
\label{sec:mode2}
Preference-based alignment was carried out with a training routine \texttt{train\_biosecurity\_dpo} using Direct Preference Optimization \citep{rafailov2023dpo}. Given each training triple $(x,\,y^+,\,y^-)$ and a reference policy $\pi_{\text{ref}}$, the DPO objective minimized
\begin{equation}
\label{eq:dpo}
\mathcal{L}_{\text{DPO}}(\theta)
=
-\,\mathbb{E}_{(x,y^+,y^-)}\;
\log \sigma\!\Big(
\beta\big[
(\log \pi_\theta(y^+\!\mid x)-\log \pi_\theta(y^-\!\mid x))
-(\log \pi_{\text{ref}}(y^+\!\mid x)-\log \pi_{\text{ref}}(y^-\!\mid x))
\big]\Big),
\end{equation}
with a default temperature $\beta=0.1$. Parameter-efficient fine-tuning was enabled via LoRA by injecting low-rank adapters into the attention and MLP projection matrices \citep{hu2022lora}, under the low-rank reparameterization
\begin{equation}
\label{eq:lora}
W_\theta \;=\; W_0 \;+\; \frac{\alpha}{r}\,B A\,,
\end{equation}
where $A\in\mathbb{R}^{r\times d_{\text{in}}}$ and $B\in\mathbb{R}^{d_{\text{out}}\times r}$ are trainable, the rank $r$ and scale $\alpha$ are hyperparameters, and $W_0$ is the original pretrained weight.

\subsection{Mode 3: Runtime alignment guard}
\label{sec:mode3}
Inputs and outputs were gated at runtime by a module \texttt{biosecurity\_alignment\_guard}. Five detection signals were aggregated. These signals were keyword hits from curated virus-related lexicons at levels $\{\texttt{L1\_custom}, \texttt{L2\_human}, \texttt{L3\_all}\}$, fuzzy string matches via RapidFuzz with partial ratio $\ge 87$ \citep{rapidfuzz}, sentence-level semantic similarity using Sentence-Transformers with cosine $\ge 0.67$ \citep{reimers2019sentencebert}, long-sequence detection for DNA or protein sequences with character length $\ge 60$, and an optional BLAST alignment screen against a curated pathogen database with identity $\ge 0.88$ and aligned length $\ge 60$ \citep{camacho2009blastplus}.  

Let $s_k(x)$ denote the score produced by signal $k$ with threshold $\tau_k$ and indicator $z_k(x)=\mathbf{1}[\,s_k(x)\ge \tau_k]$. The guard decision followed a lexicographic priority
\begin{equation}
\label{eq:guard}
d(x)=
\begin{cases}
\textsf{block}_{\text{blast}}, & z_{\text{blast}}=1,\\
\textsf{block}_{\text{longseq}}, & z_{\text{blast}}=0,\ z_{\text{longseq}}=1,\\
\textsf{block}_{\text{semantic}}, & z_{\text{blast}}=0,\ z_{\text{longseq}}=0,\ z_{\text{semantic}}=1,\\
\textsf{block}_{\text{fuzzy}}, & z_{\text{blast}}=0,\ z_{\text{longseq}}=0,\ z_{\text{semantic}}=0,\ z_{\text{fuzzy}}=1,\\
\textsf{block}_{\text{keyword}}, & z_{\text{blast}}=0,\,\dots,\,z_{\text{fuzzy}}=0,\ z_{\text{keyword}}=1,\\
\textsf{allow/warn}, & \text{otherwise.}
\end{cases}
\end{equation}
This ordering places biologically specific and high-precision checks first and lexical pattern matches later. The arrangement increases robustness to prompt variants and simplifies operating-point calibration when the thresholds $\{\tau_k\}$ are tuned under an FPR budget. In practice, the thresholds were selected on a held-out validation set to minimize the jailbreak success rate subject to $\mathrm{FPR}\le\epsilon$. The same policy was applied at both the \emph{pre} and \emph{post} guard positions.

Under default setting, evaluations used the input guard in strict mode with keyword and fuzzy matching enabled and with semantic matching and BLAST disabled. Ablation tests with semantic matching and BLAST enabled are reported where relevant.

\subsection{Mode 4: Automated red-team evaluation}
\label{sec:mode4}
Automated red teaming was conducted by a specialized attacker agent in a repeating three-stage loop. The stages were a strict pre-generation guard without BLAST, the model $M_\theta$ generating a continuation, and a strict post-generation guard with BLAST. The attacker iteratively mutated prompts based on prior outcomes. Any successful prompt that bypassed defenses was recycled into \textbf{Mode~3} as a new guard rule or threshold adjustment and into \textbf{Mode~2} as a new preference pair for alignment training \citep{zhou2025autoredteamer}. For guard operating-point selection, detection thresholds $\lambda$ were chosen to minimize the jailbreak success rate subject to a false-positive budget
\begin{equation}
\label{eq:op}
\min_{\lambda}\ \text{JSR}(\lambda)\quad \text{s.t.}\ \text{FPR}(\lambda)\le \epsilon\,,
\end{equation}
and ROC and precision–recall curves were obtained by sweeping $\lambda$ over the signal thresholds in Eq.~\eqref{eq:guard}.

\FloatBarrier

\section{Results}
\label{sec:results}

\subsection{Mode 1: Dataset sanitisation on CORD-19}
The CORD-19 corpus \citep{wang2020cord19}, a benchmark dataset of biomedical research articles, was used to evaluate Mode~1 dataset sanitisation. Sanitisation was applied at three keyword strictness levels. The removal rate increased monotonically with stricter filtering. The Level~2 configuration pruned roughly one-fifth of the corpus while preserving about 80\% of the entries, whereas the Level~3 configuration removed the majority of entries. This demonstrates a safety–utility trade-off as filtering becomes more aggressive.
\begin{figure}[!hb]
  \centering
  \includegraphics[width=0.62\linewidth,keepaspectratio,
                   trim=10pt 6pt 10pt 2pt,clip]{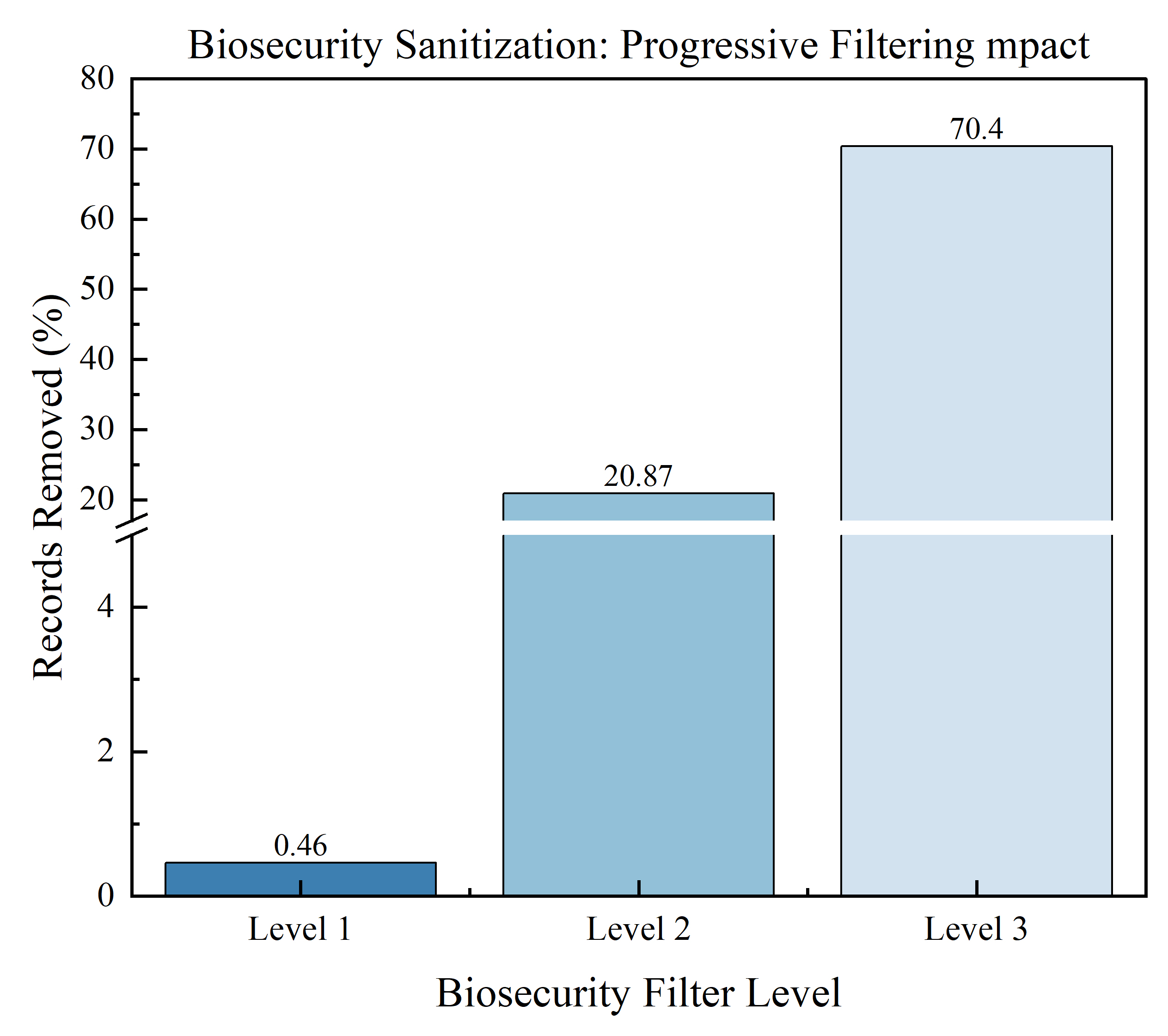}
  \caption{Removal rate on CORD-19 at each biosecurity level.
  A monotonic increase is observed from Level~1 to Level~3.
  The Level~2 configuration removes about 21\% of entries,
  whereas the Level~3 configuration removes about 70\%.}
  \label{fig:mode1-removal-rate}
\end{figure}
\FloatBarrier

\subsection{Mode 2: Safety alignment via DPO}
\begin{figure}[!t]
  \centering
  \includegraphics[width=0.55\linewidth,keepaspectratio,
                   trim=10pt 8pt 10pt 6pt,clip]{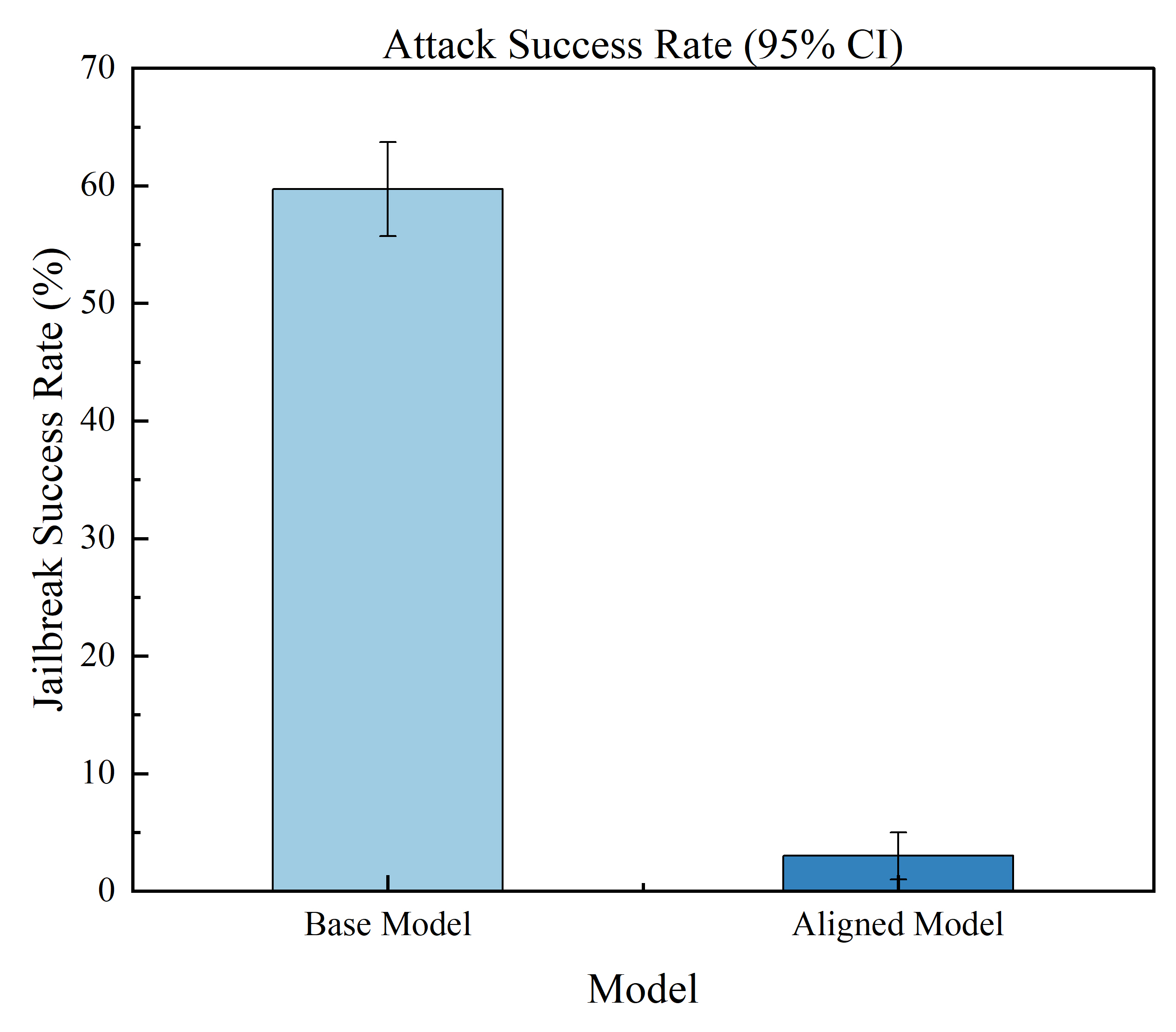}
  \caption{\textbf{Mode 2 — Attack success rate with 95\% confidence intervals.}
  Preference alignment (DPO+LoRA) lowers ASR from 59.7\% to 3.0\% on the expanded
  adversarial set. Error bars indicate Clopper–Pearson 95\% confidence intervals.}
  \label{fig:mode2-asr-ci}
\end{figure}

Direct Preference Optimization with LoRA was applied to align the base model toward refusals and safe completions. On a representative 60-prompt evaluation set, the jailbreak success rate decreased from 30\% to 10\%. The safe-accept rate increased from 70\% to 90\% with the pre-guard block rate fixed at 30\%. On an expanded adversarial set, the end-to-end ASR decreased from 59.7\% (95\% CI 55.6–63.7) to 3.0\% (1.0–5.0), meeting the below 5\% target (Fig.~\ref{fig:mode2-asr-ci}).
\FloatBarrier

\subsection{Mode 3: Runtime alignment guard}
\begin{figure}[!t]
  \centering
  \includegraphics[width=0.95\linewidth,keepaspectratio,trim=2pt 2pt 2pt 2pt,clip]{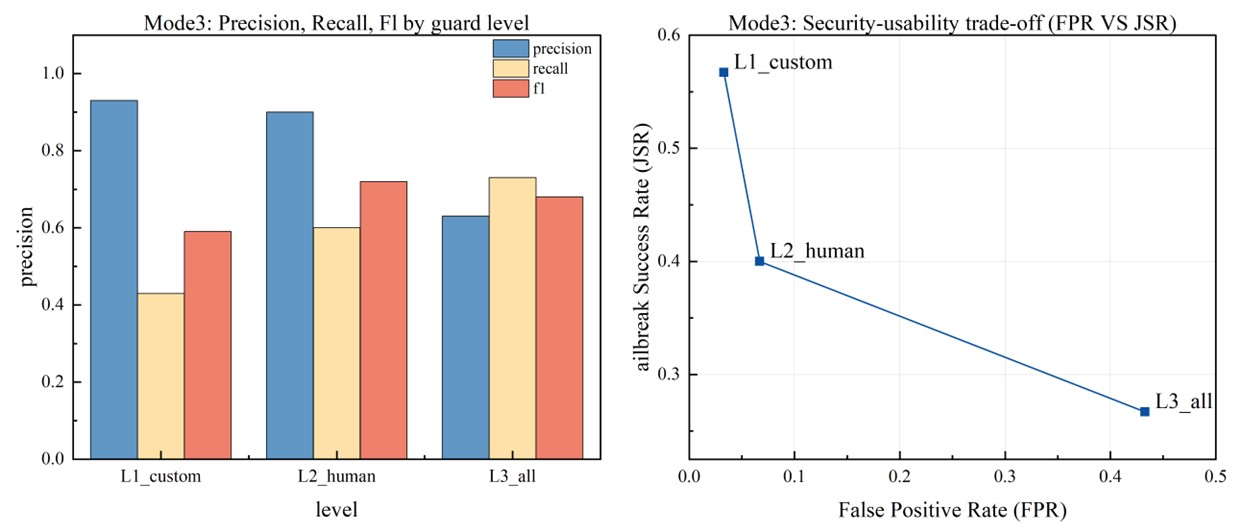}
  \caption{\textbf{Mode 3. Guard performance and the security–usability trade-off.}
  Top panel shows precision, recall, and F1 for guard configurations (L1\_custom, L2\_human, L3\_all).
  Bottom panel shows FPR versus JSR, where the lower-left region is preferred.
  The \texttt{L2\_human} configuration attains the highest F1 at a low FPR.
  The \texttt{L3\_all} configuration achieves the lowest JSR at the cost of a higher FPR.}
  \label{fig:mode3-metrics}
\end{figure}

\begin{figure}[!t]
  \centering
  \includegraphics[width=0.7\linewidth,trim=2pt 2pt 2pt 2pt,clip]{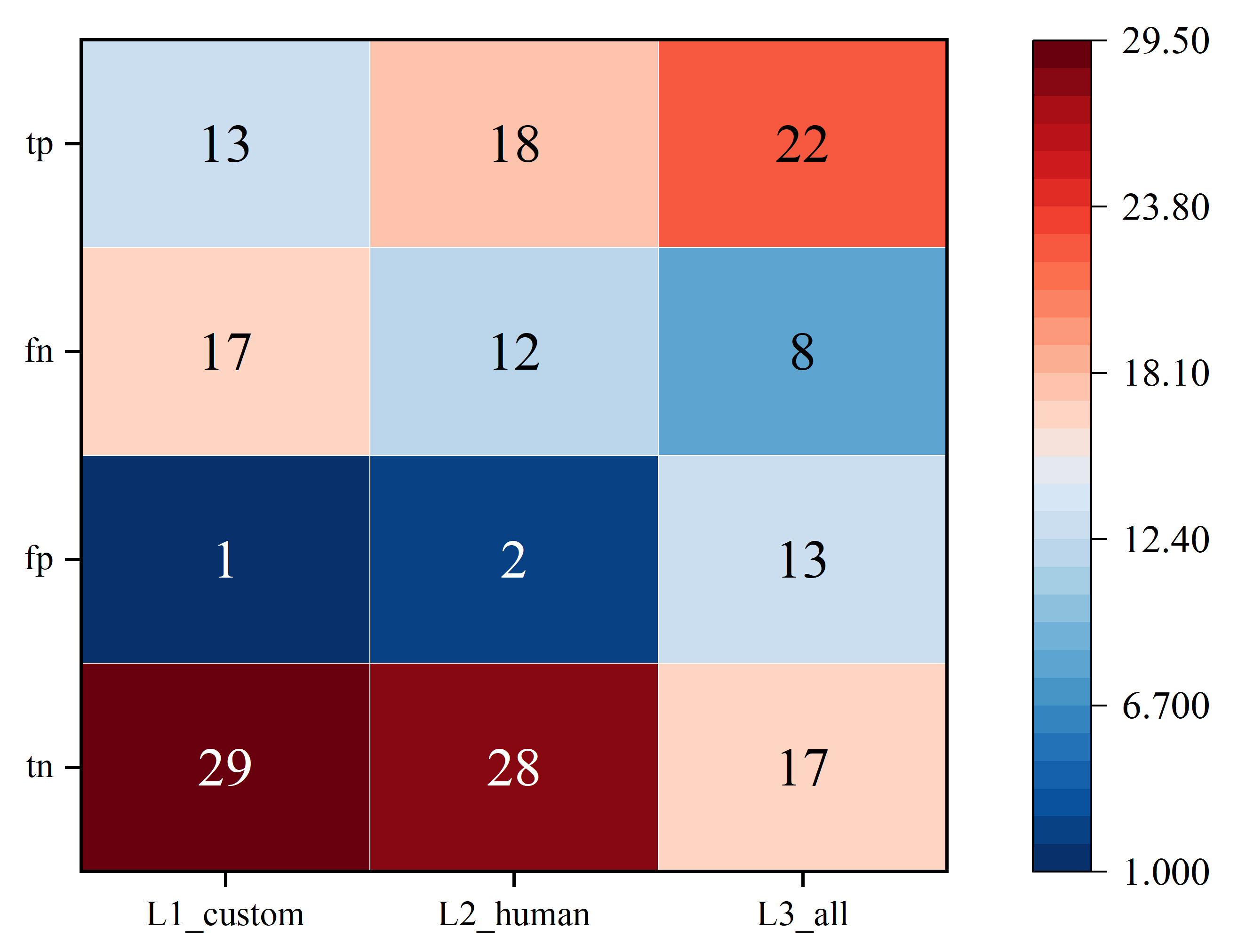}
  \caption{\textbf{Mode 3. Confusion outcomes across guard levels.}
  The heatmap merges the three guards into one matrix.
  Rows list outcomes \textit{tp}, \textit{fn}, \textit{fp}, \textit{tn}.
  Columns correspond to \textit{L1\_custom}, \textit{L2\_human}, \textit{L3\_all}.
  Each cell gives the count on the 60-prompt evaluation set.
  From L1 to L3, true positives increase and false negatives decrease.
  False positives rise and true negatives fall.
  The pattern reflects increasing strictness and the expected security–usability trade-off.}
  \label{fig:mode3-confusions}
\end{figure}

The guard was evaluated on a labeled set of 60 prompts under three keyword strictness levels. A clear security–usability trade-off was observed (Figs.~\ref{fig:mode3-metrics}--\ref{fig:mode3-confusions}). The \textbf{L1\_custom} configuration achieved the lowest FPR (0.033) but exhibited the highest JSR (0.567). The \textbf{L2\_human} configuration attained the highest F1 (0.720) with precision 0.900 and recall 0.600 at an FPR of 0.067. The \textbf{L3\_all} configuration yielded the lowest JSR (0.267) and the highest recall (0.733) but incurred an FPR of 0.433 and a reduced precision of 0.629.

\FloatBarrier

\subsection{Mode 4: Automated red-team evaluation}
End-to-end stress testing with adaptive adversarial prompts was conducted to assess post-alignment robustness under distribution shift. Improvements were observed across the board. Mean precision increased from $0.752\pm0.010$ to $0.868\pm0.005$, mean recall increased from $0.674\pm0.033$ to $0.910\pm0.017$, and mean FPR decreased from $0.268\pm0.025$ to $0.027\pm0.012$. The allocation of defensive actions shifted upstream. The pre-guard block rate increased from 15\% to 40\%, while the post-guard block rate decreased from 25\% to 5\%. The safe-completion rate remained stable at approximately 55--60\%.

\begin{figure}[!hb]
  \centering
  \includegraphics[width=0.95\linewidth,keepaspectratio]{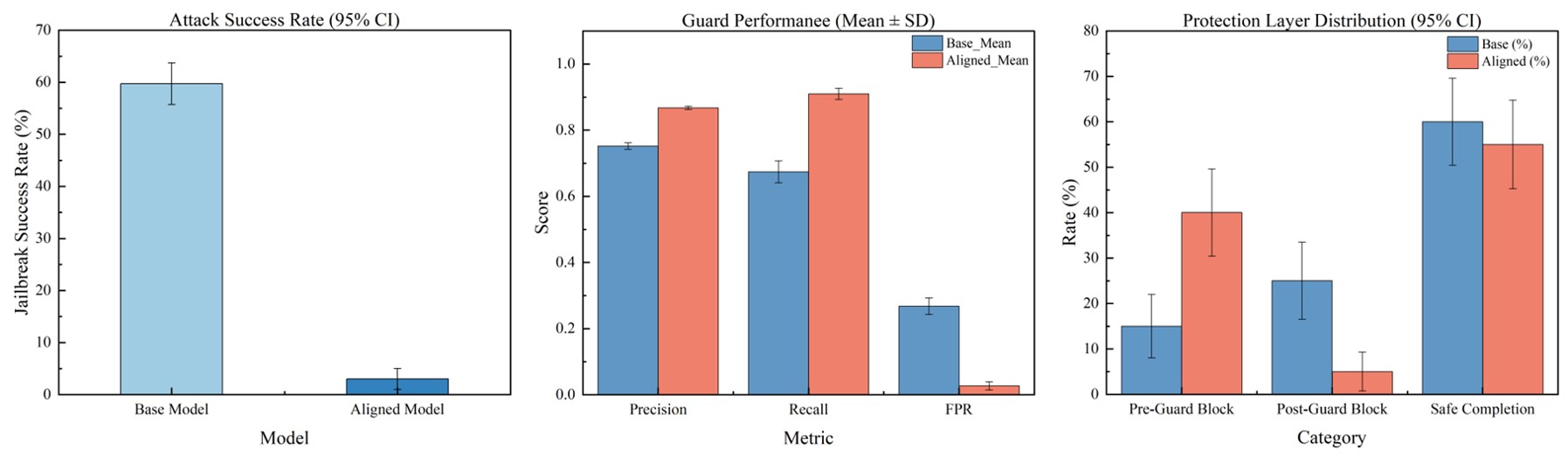}
  \caption{\textbf{Mode 4. Automated red-team evaluation (aggregated results).}
  Left panel shows ASR with 95\% confidence intervals and indicates a reduction from 59.7\% to 3.0\% after alignment.
  Middle panel shows guard performance with mean and standard error where precision and recall increase and FPR drops after alignment.
  Right panel shows the distribution of protection across guard layers with 95\% confidence intervals over 100 runs and highlights a shift toward upstream blocking while maintaining a stable safe-completion rate.}
  \label{fig:mode4-panels}
\end{figure}
\FloatBarrier

\section{Discussion}

\paragraph{Summary of findings.}
In this research, the pipeline meets a predefined end-to-end safety target while preserving benign utility across the four lifecycle modes. Training-time alignment delivers the largest single gain and reduces reliance on inference-time filtering. The guard admits calibrated operating points under an explicit false-positive budget. Tiered sanitization provides source-level control with a measurable safety–utility trade-off. Continuous automated red teaming improves robustness and shifts protection toward the pre-guard stage.

\paragraph{Upstream data sanitization and the safety–utility trade-off (Mode~1).}
Tiered filtering on CORD-19 shows a monotonic relation between strictness and removal, with a mid-level list retaining most of the corpus while excising domain-specific risk content. This supports source-level governance as a complement to alignment and guarding and aligns with sector guidance and screening protocols \citep{who2022lifesciences,nsabb2023framework,igsc2024hsp}. The choice of level should reflect acceptable loss of coverage for downstream tasks and the requirement to limit exposure to dual-use material.

\paragraph{Training-time alignment (Mode~2).}
Preference-based alignment with DPO and parameter-efficient adapters establishes a strong safety prior in the base policy. End-to-end evaluation indicates that harmful completions fall below a prespecified target while helpful responses remain accessible. The pattern is consistent with alignment evidence that preference optimization and constitution-guided critique reduce disallowed behaviors with modest annotation burden \citep{ouyang2022instructgpt,rafailov2023dpo,bai2022constitutional,hu2022lora}. In operation, alignment lowers the load on inference-time filtering and narrows the set of prompts that require strict post hoc intervention. This accords with governance guidance that emphasizes pre-deployment assurance for high-stakes applications \citep{whitehouse2023eo14110,eurlex2024aiact,nist2023airmf}.

\paragraph{Inference-time guardrails and operating-point selection (Mode~3).}
The guard exhibits the expected precision–recall trade-off across keyword strictness levels, with a mid-level configuration balancing true blocking and user-facing false positives. Because multiple biology-aware signals are composed by specificity and robustness, thresholds can be tuned under an explicit false-positive budget as formalized in the operating-point objective. Quantitative calibration is supported by ROC and precision–recall analyses \citep{fawcett2006roc,davis2006prroc}. Compared with single-method detectors such as standalone safety classifiers or purely randomized defenses, a composite guard enables principled control of sensitivity while preserving utility \citep{llamaguard2023,shankar2024guardrails,robey2023smoothllm}.

\paragraph{Continuous automated red teaming (Mode~4).}
A closed-loop attacker–defender process improves robustness over iterations. After adaptive testing and feedback, a larger share of threats is intercepted at input and fewer reach post-generation review while benign completions remain stable. This observation aligns with reports that iterative evaluation and integration of discovered attacks strengthen robust refusal and reduce regressions \citep{mazeika2024harmbench,chao2024jailbreakbench}. It also accords with findings that autonomous red teams expand coverage beyond static benchmarks and sustain discovery over time \citep{zhou2025autoredteamer,ganguli2022redteaming}. In deployment, such a loop offers an auditable mechanism for continuous hardening that updates guard rules and preference data as the threat surface evolves.

\paragraph{Limitations and outlook.}
The present scope concerns text-only assistance. Sequence-generating models for DNA or proteins and multimodal lab-control settings are not evaluated. Some experiments rely on a compact challenge set, which introduces statistical uncertainty even with exact binomial intervals. Independent replication on public suites such as JailbreakBench and HarmBench would further substantiate generalization \citep{chao2024jailbreakbench,mazeika2024harmbench}. Future extensions include learned detectors within the guard stack, adaptive thresholding to reduce false positives under stringent budgets, and expanded red-team agents that target dialog-level strategies and multilingual prompts \citep{liu2024autodan,li2024wildjailbreak,zou2023universal}. These directions aim to keep a calibrated operating point as models and adversaries co-evolve.

\bibliography{iclr2025_conference}
\bibliographystyle{iclr2025_conference}

\end{document}